\documentclass[prl,showpacs,superscriptaddress,twocolumn]{revtex4-1}
\usepackage{graphicx}		
\usepackage{dcolumn}		
\usepackage{bm}					
\usepackage{textcomp}		

\begin{document}
\title{Erratum : Enhancement of the Spin Accumulation at the Interface Between a Magnetic Tunnel Junction and a Semiconductor.\\ {[}\textbf{Phys. Rev. Lett.~\textbf{102}, 036601 (2009)}]}
\author{M. Tran}
\author{H. Jaffr\`es}
\author{J.-M. George} \email{jean-marie.george@thalesgroup.com}
\author{A. Fert}
\author{A. Miard}
\author{A. Lema\^{\i}tre}

\date{\today}
\maketitle

In the beginning of a previous letter~\cite{tran2009}, we have given a litteral expression for a characteristic threshold resistance $r_1$ of a semiconductor~\cite{ieeefert2007} at an interface with a magnetic tunnel junction as $r_1=(\rho l_{sf})~W/w$ where $\rho$ is the semiconductor resistivity, $l_{sf}$ its spin diffusion length, $W$ the contact width and $w$ the channel thickness. This was derived in the standard theory of spin diffusion/relaxation in the conduction band~\cite{valetfert1993} taking into account a geometric renormalization when the spin transport is made along a lateral channel~\cite{fertjaffres2001}. However, the previous expression is correct only in the limit where $l_{sf}\gg (W,~w)$. In the case of a spin diffusion length well shorter than the contact width $W$ ($l_{sf}\ll W$) but still in the limit $l_{sf}\gg w$ like in the experimental situation, $r_1$ should instead write $r_1=\rho~(l_{sf})^2/w$~\cite{note1}. In this limit, it results that the correct spin resistance area product writes $R_S.A=\gamma^2 r_1=\gamma^2 \rho l_{sf}^2/w\simeq \rho l_{sf}^2/w \simeq 200~\Omega.~\mu m^2$ for a maximal value of $\gamma=1$ ($\gamma$ is the tunnel spin asymmetry coefficient~\cite{note2}) and a maximum value of $l_{sf}=1~\mu$m for the range of doping used~\cite{dzhioev2002}. Since the expected $R_S.A$ product appears even more smaller than the one previously reported (1 $k\Omega.~\mu m^2$), it does not change the core of the present letter emphasizing on the role of the surface states (or localized states) at the direct oxide/semiconductor interfaces giving rise to a strong amplification of the spin signal at the level of 3.6~$M\Omega.~\mu m^2$ or beyond~\cite{dash2011} as experimentally observed.

\end {document}